\documentclass[a4paper,twoside,12pt]{article}
\input{obsjkt.sty}

\newcommand{\Msun}{\ensuremath{~{\rm M}_\odot}}                   
\newcommand{\Rsun}{\ensuremath{~{\rm R}_\odot}}                   
\newcommand{\rhosun}{\ensuremath{~\rho_\odot}}                    
\newcommand{\Teff}{\ensuremath{T_{\rm eff}}}                      
\newcommand{\FeH}{\ensuremath{\rm [Fe/H]}}                        
\newcommand{\EBV}{\ensuremath{E(B\!-\!V)}}                        
\newcommand{\Grp}{\ensuremath{G_{\rm RP}}}                        
\newcommand{\degr}{\ensuremath{^\circ}}                           
\renewcommand{\kms}{~km~s$^{-1}$}                                 

\newcommand{\corot}{\textit{CoRoT}}
\newcommand{\kepler}{\textit{Kepler}}
\newcommand{\hip}{\textit{Hipparcos}}
\newcommand{\gaia}{\textit{Gaia}}
\newcommand{\targ}{V454~Aur}
\newcommand{\targfull}{V454~Aurigae}

\newcommand{\Msunnom}{\hbox{$\mathcal{M}^{\rm N}_\odot$}}
\newcommand{\Rsunnom}{\hbox{$\mathcal{R}^{\rm N}_\odot$}}
\newcommand{\Lsunnom}{\hbox{$\mathcal{L}^{\rm N}_\odot$}}

\usepackage{rotating}

\begin{document} 

\OBSheader{Rediscussion of eclipsing binaries: \targ}{J.\ Southworth}{2024 August}

\OBStitle{Rediscussion of eclipsing binaries. Paper XIX. \\ The long-period solar-type system V454 Aurigae}

\OBSauth{John Southworth}

\OBSinstone{Astrophysics Group, Keele University, Staffordshire, ST5 5BG, UK}


\OBSabstract{\targ\ is an eclipsing binary system containing two solar-type stars on an orbit of relatively long period ($P = 27.02$~d) and large eccentricity ($e = 0.381$). Eclipses were detected using data from the \hip\ satellite, and a high-quality double-lined spectroscopic orbit has been presented by Griffin \cite{Griffin01obs}. The NASA Transiting Exoplanet Survey Satellite (TESS) has observed the system during eight sectors, capturing ten eclipses in their entirety. \targ\ is unusual in that the primary star -- the star eclipsed at the deeper minimum -- is less massive, smaller \emph{and} cooler than its companion. This phenomenon can occur in certain configurations of eccentric orbits when the stars are closer together at the primary eclipse, causing a larger area to be eclipsed than at the secondary. We use the radial velocity measurements from Griffin and the light curves from TESS to determine the masses and radii of the component stars for the first time, finding masses of $1.034 \pm 0.006$\Msun\ and $1.161 \pm 0.008$\Msun, and radii of $0.979 \pm 0.003$\Rsun\ and $1.211 \pm 0.003$\Rsun. Our measurement of the distance to the system is consistent with that from the \gaia\ DR3 parallax. A detailed spectroscopic study to determine chemical abundances and more precise temperatures is encouraged. Finally, we present equations to derive the effective temperatures of the stars from the inferred temperature of the system as a whole, plus the ratio of the radii and either the surface brightness or light ratio of the stars.}


\section*{Introduction}

Detached eclipsing binaries (dEBs) are our primary source of directly measured masses and radii of normal stars \cite{Andersen91aarv,Torres++10aarv,Me15debcat}, obtained from the analysis of time-series photometry and radial velocity (RV) measurements. Early studies of these objects were hampered by the difficulty of obtaining high-quality photometry covering all orbital phases \cite{Stebbins10apj,Stebbins11apj}, particularly with the equipment in use at the time \cite{Stebbins16obs,Wesselink41anlei,Gaposchkin53anhar}. 

Improvements required the availability of extensive observing time on small telescopes (e.g.\ refs.~\citenum{Kron42apj,Delandtsheer83aas}), preferably operated robotically (e.g.\ refs.~\citenum{Gronbech++76aas,Florentinnielsen++87msngr}). The operation of an increasing number of small survey instruments for stellar variability (e.g.\ ref.~\citenum{Pojmanski97aca}) or planetary transits \citep{Bakos+02pasp,Pepper+07pasp,Pollacco+06pasp} has resulted in the acquisition of extensive photometry for millions of bright stars. Some of these targets are dEBs for which precise radii could be obtained \cite{Helminiak+09mn,Me+11mn2}. 

The recent generation of space telescopes dedicated to the detection of transiting extrasolar planets from time-series survey photometry -- such as \corot\ \cite{Auvergne+09aa}, \kepler\ \cite{Borucki16rpph} and TESS (the Transiting Exoplanet Survey Satellite \cite{Ricker+15jatis}) -- has hugely increased the extent and precision of photometric archives \cite{Me21univ}. This has led to a fundamental change in the number of dEBs both known \cite{Deleuil+08aa,Kirk+16aj,Prsa+21apjs} and analysed in detail \cite{Bass+12apj,Me20obs,Themessl+18mn}.

It is more difficult to obtain good observational datasets for dEBs with longer orbital periods ($P$). On the spectroscopic side, the velocity amplitudes scale according to $P^{-1/3}$ so the size of the measurable signal decreases. On the photometric side, the eclipses become longer and rarer and thus poorly suited to ground-based observation. However, as $P$ increases, photometric study gets harder more quickly than spectroscopic study due to the more time-critical nature of the required observations. The result is that extensive sets of RVs have been obtained for some longer-period dEBs without the accompanying photometric analysis needed for the determination of the full physical properties of the component stars. This was the situation for \targ, except that high-quality photometry is now available from TESS. The current work presents an analysis of these new data and the first measurement of precise masses and radii of its constituents.


\section*{\targfull}

\begin{table}[t]
\caption{\em Basic information on \targfull. \label{tab:info}}
\centering
\begin{tabular}{lll}
{\em Property}                            & {\em Value}                 & {\em Reference}                      \\[3pt]
Right ascension (J2000)                   & 06:22:03.06                 & \citenum{Gaia21aa}                   \\
Declination (J2000)                       & +34:35:50.5                 & \citenum{Gaia21aa}                   \\
Henry Draper designation                  & HD 44192                    & \citenum{CannonPickering18anhar2}    \\
\textit{Hipparcos} designation            & HIP 30270                   & \citenum{Hipparcos97}                \\
\textit{Gaia} DR3 designation             & 3440630513359154688         & \citenum{Gaia21aa}                   \\
\textit{Gaia} DR3 parallax                & $15.3669 \pm 0.0217$ mas    & \citenum{Gaia21aa}                   \\          
TESS\ Input Catalog designation           & TIC 138333980               & \citenum{Stassun+19aj}               \\
$B$ magnitude                             & $8.22 \pm 0.03$             & \citenum{Hog+00aa}                   \\          
$V$ magnitude                             & $7.65 \pm 0.01$             & \citenum{Hog+00aa,Olsen83aas}        \\          
$J$ magnitude                             & $6.589 \pm 0.023$           & \citenum{Cutri+03book}               \\
$H$ magnitude                             & $6.374 \pm 0.027$           & \citenum{Cutri+03book}               \\
$K_s$ magnitude                           & $6.297 \pm 0.018$           & \citenum{Cutri+03book}               \\
Spectral type                             & F8~V + G1-2~V               & \citenum{Griffin01obs}               \\[3pt]
\end{tabular}
\end{table}

\targ\ (Table~\ref{tab:info}) was found to be eclipsing using data from the \hip\ satellite \cite{Hipparcos97} and given its variable star name by Kazarovets et al.\ \cite{Kazarovets+99ibvs}. The object was subsequently observed by Griffin \cite{Griffin01obs} (hereafter G01) in Paper 160 of his \textit{Spectroscopic Binary Orbits from Photoelectric Radial Velocities} series, alongside V455~Aur (since revisited by Southworth \cite{Me21obs4}) and UW~LMi (reanalysed by Graczyk et al.\ \cite{Graczyk+22aa}).


G01 originally added \targ\ to his observing list based on its overluminosity (from the \hip\ parallax) being an indication of binarity \cite{SuchkovMcmaster99apj}. He corrected the original suggested period of 3.2057~d to its true value of 27.02~d using a set of 62 spectral cross-correlation functions \cite{Griffin67apj} observed over a time interval of 386~d. The substantial orbital eccentricity means the RVs of the stars are significantly different at times of eclipse, a fact used by G03 to confirm the presence of both primary and secondary eclipses by the weakening of the dip of a given star in the cross-correlation functions\footnote{despite noting a ``complete lack of meteorological co\"operation'' for some of these observations} (see his fig.~2). 

G01 estimated a light ratio of approximately 0.58 from the ratio of the cross-correlation dips: this should be interpreted as the ratio of spectral line strengths in the wavelength interval close to the peak of the Johnson $B$ band. From this and the colour indices of the system, he inferred spectral types of F8~V and G1-2~V.

The only other published information worth mentioning at this point are measurements of the effective temperature (\Teff) and iron abundance (\FeH) of the system. Both come from the Geneva-Copenhagen Survey \cite{Nordstrom+04aa}, and are $\Teff = 6064 \pm 80$~K and $\FeH = -0.08$ (Casagrande et al.\ \cite{Casagrande+11aa}) and $\Teff = 6030$~K and $\FeH = -0.14$ (Holmberg et al.\ \cite{Holmberg+07aa}).

The $BV$ magnitudes in Table~\ref{tab:info} come from the Tycho experiment \cite{Hog+00aa} on the \hip\ satellite. Each comprise the average of 85 measurements, well-distributed in orbital phase and with only a few obtained during an eclipse. The $JHK_s$ magnitudes are from 2MASS \cite{Cutri+03book} and were obtained at a single epoch corresponding to orbital phase 0.679, which is not within an eclipse.



\section*{Photometric observations}

\begin{figure}[t] \centering \includegraphics[width=\textwidth]{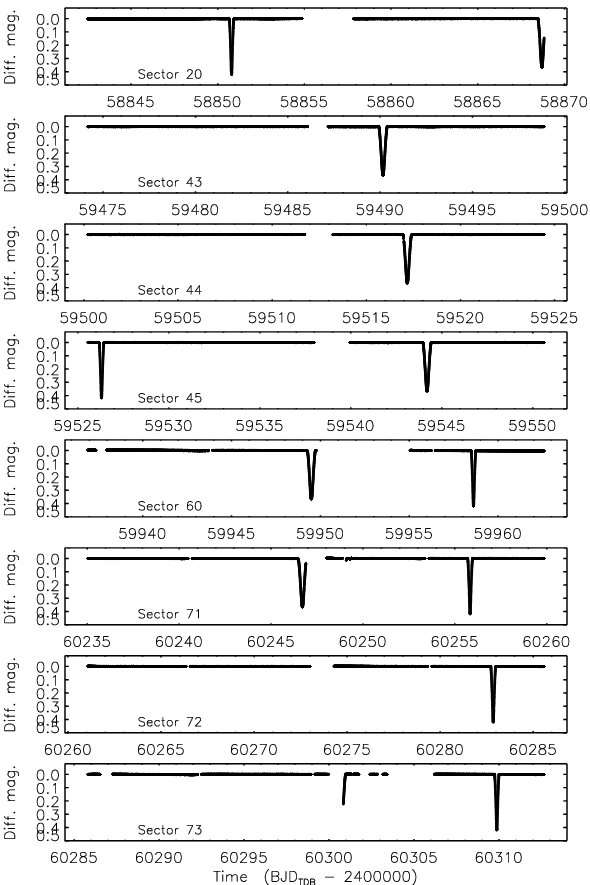} \\
\caption{\label{fig:time} TESS\ short-cadence SAP photometry of \targ. The flux 
measurements have been converted to magnitude units then rectified to zero magnitude 
by subtraction of the median. Each panel shows one TESS sector (labelled).} \end{figure}

%

\targ\ has been observed in eight sectors by TESS \cite{Ricker+15jatis}, beginning with sector 20 (2020 January) and ending in sector 73 (2023 December). Data in all sectors were obtained at a cadence of 120~s as well as other cadences including 20~s, 600~s and 1800~s. We downloaded all 120~s cadence data from the NASA Mikulski Archive for Space Telescopes (MAST\footnote{\texttt{https://mast.stsci.edu/portal/Mashup/Clients/Mast/Portal.html}}) using the {\sc lightkurve} package \cite{Lightkurve18}. We adopted the simple aperture photometry (SAP) data from the TESS-SPOC data reduction \cite{Jenkins+16spie} with a quality flag of ``hard''. These were normalised using {\sc lightkurve} and converted to differential magnitude.

The resulting light curves are shown in Fig.~\ref{fig:time}, divided according to sector. It can be seen that six primary eclipses were observed, and all are fully covered. There are also seven secondary eclipses, but only four are fully covered by the available observations. The eccentric nature of the system is clear from the facts that the secondary eclipses are longer than the primary eclipses, and they do not occur midway between successive primary eclipses. In the following analyses we adopt the standard approach of labelling the deeper eclipse as the primary eclipse, the primary star (star~A) as the star eclipsed in primary eclipse, and the secondary star as star~B.

\clearpage

We queried the \gaia\ DR3 database\footnote{\texttt{https://vizier.cds.unistra.fr/viz-bin/VizieR-3?-source=I/355/gaiadr3}} and found a total of 82 objects within 2~arcmin of \targ. Of these, the brightest is fainter than our target by 5.94~mag in the \Grp\ band, and the brightest star within 1~arcmin is fainter by 7.77~mag in \Grp. We therefore expect the TESS light curve to suffer from contamination at a level below 1\%.


\section*{Light curve analysis}

\targ\ contains two well-detached stars and the TESS data are extensive, so the system is well suited to analysis with the {\sc jktebop}\footnote{\texttt{http://www.astro.keele.ac.uk/jkt/codes/jktebop.html}} code \cite{Me++04mn2,Me13aa} (version 43). Fitting the full 128\,226 datapoints simultaneously took a significant amount of computing time, which could be avoided by rejecting data away from eclipse and thus of negligible information content. We therefore extracted the ten fully-observed eclipses from the light curve, including 0.3~d (primary) and 0.45~d (secondary) of data outside eclipse, giving a more tractable 5053 datapoints for detailed analysis. 

The fitted parameters were the fractional radii of the stars ($r_{\rm A}$ and $r_{\rm B}$), expressed as their sum ($r_{\rm A}+r_{\rm B}$) and ratio ($k = {r_{\rm B}}/{r_{\rm A}}$), the central surface brightness ratio ($J$), the orbital inclination ($i$) and period ($P$), and a reference time of primary minimum ($T_0$). Orbital eccentricity ($e$) and the argument of periastron ($\omega$) were fitted via the combinations $e\cos\omega$ and $e\sin\omega$. We also fitted for a second-order polynomial brightness variation for each eclipse to remove any remaining slow changes in brightness due to either instrumental or astrophysical effects.

Limb darkening was included using the power-2 law \cite{Hestroffer97aa,Me23obs2} with the linear coefficient ($c$) fitted and the power coefficient ($\alpha$) fixed to a theoretical value \cite{ ClaretSouthworth22aa,ClaretSouthworth23aa}. The two stars have sufficiently similar limb darkening characteristics that we assumed the same coefficients for both. We initially included third light as a fitted parameter, but found that it always became small and insignificant. We therefore fixed it at zero for our definitive solution, which is given in Table~\ref{tab:jktebop}.

Our initial solutions of the light curve with reasonable estimates of the starting parameters yielded an unexpected outcome. The particular geometry of the orbit of \targ\ means that the stars are significantly closer to each other during primary than secondary eclipse, which combined with $i < 90\degr$ means less of the stars are eclipsed during secondary than primary. The only way to get the secondary eclipse deep enough to match the data is for star~B to be both larger \emph{and} hotter than star~A. In support of this counterintuitive result is that the value of $\omega$ from the light curve differs by 180\deg\ from the spectroscopic one (G01). We confirmed it by fitting for the RVs from G01 simultaneously with the TESS light curve and finding that our identifications of the stars are swapped relative to Griffin's. \targ\ is therefore a rare example of a dEB where the secondary star is larger, hotter and more massive than the primary. This can only occur for specific $\omega$ values in an eccentric orbit.


\begin{figure}[t] \centering \includegraphics[width=\textwidth]{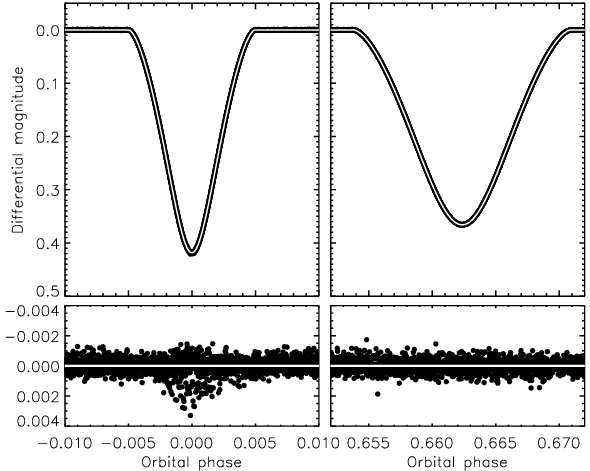} \\
\caption{\label{fig:phase} {\sc jktebop} best fit to the 120-s cadence TESS light curves 
of \targ. The data are shown with filled circles and the best fit with a white-on-black 
line. The residuals are shown on an enlarged scale in the lower panel.} \end{figure}

\begin{figure}[t] \centering \includegraphics[width=\textwidth]{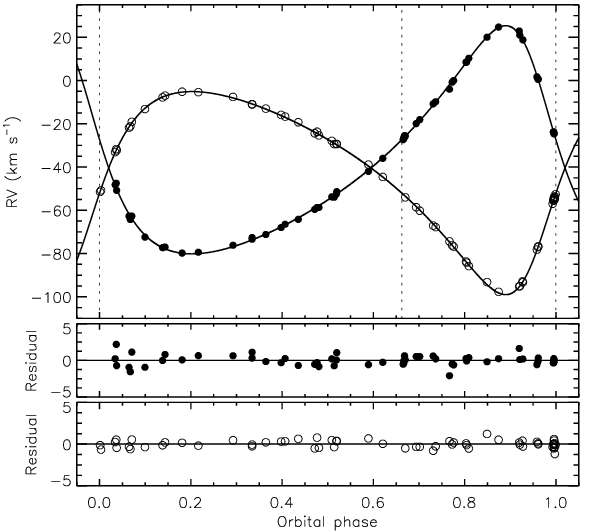} \\
\caption{\label{fig:rv} RVs of \targ\ from G01 (filled circles for star~A and 
open circles for star~B), compared to the best fit from {\sc jktebop} (solid 
lines). The times of eclipse are given using vertical dotted lines. The residuals 
are given in the lower panels separately for the two components.} \end{figure}

Once a suitable solution was established, we ran  Monte Carlo and residual-permutation solutions \citep{Me08mn} to obtain reliable errorbars \citep{Me21obs4}. We fitted both the TESS and RV data, allowing a separate systemic velocity for the two stars. The best fits are given in Fig.~\ref{fig:phase} for the light curve and Fig.~\ref{fig:rv} for the RV curves, and the properties are collected in Table~\ref{tab:jktebop}. The solution is extremely well-determined, with uncertainties in the fractional radii of 0.2\% for star~A and 0.04\% for star~B. We imposed a minimum uncertainty of 0.2\% on the fractional radii based on the independent-analyses tests described by Maxted et al.\ \cite{Maxted+20mn} for the similar system AI~Phoenicis. Our results are in good agreement with those of G01, after accounting for his different choice of primary star. The light ratio also agrees with the value found by G01; it should be remembered that the wavelength interval observed by G01 is significantly bluer than the TESS passband.

\begin{table} \centering
\caption{\em \label{tab:jktebop} Parameters of \targ, with their 1$\sigma$ uncertainties, 
measured from the TESS sector 55 light curves using the {\sc jktebop} code.}
\setlength{\tabcolsep}{4pt}
\begin{tabular}{lcc}
{\em Parameter}                           &              {\em Value}            \\[3pt]
{\it Fitted parameters:} \\                                                   
Primary eclipse time (BJD$_{\rm TDB}$)    & $ 2459526.296873   \pm 0.000015   $ \\
Orbital period (d)                        & $      27.0198177  \pm 0.00000082 $ \\
Orbital inclination (\degr)               & $      89.2084     \pm  0.0023    $ \\
Sum of the fractional radii               & $       0.044456   \pm  0.00033   $ \\
Ratio of the radii                        & $       1.2368     \pm  0.0027    $ \\
Central surface brightness ratio          & $       1.2059     \pm  0.0020    $ \\
LD coefficient $c$                        & $       0.623      \pm  0.010     $ \\
LD coefficient $\alpha$                   &            0.545 (fixed)            \\
$e\cos\omega$                             & $       0.246836 \pm  0.000018    $ \\
$e\sin\omega$                             & $       0.28965  \pm  0.00023     $ \\
Velocity amplitude of star~A (\kms)       & $      52.75     \pm  0.17        $ \\
Velocity amplitude of star~B (\kms)       & $      46.95     \pm  0.11        $ \\
Systemic velocity of star~A (\kms)        & $     -40.41     \pm  0.02        $ \\ 
Systemic velocity of star~B (\kms)        & $     -40.48     \pm  0.02        $ \\ 
{\it Derived parameters:} \\                                                   
Fractional radius of star~A               & $       0.019875   \pm  0.000039  $ \\
Fractional radius of star~B               & $       0.024582   \pm  0.000010  $ \\
Light ratio $\ell_{\rm B}/\ell_{\rm A}$   & $       1.8448     \pm  0.0053    $ \\[3pt]
Orbital eccentricity                      & $       0.38056    \pm  0.00017   $ \\
Argument of periastron ($^\circ$)         & $      49.562      \pm  0.025     $ \\
\end{tabular}
\end{table}


\section*{Physical properties and distance to \targ}

We determined the physical properties of the \targ\ system using the results in Table~\ref{tab:jktebop} from the {\sc jktebop} analysis. We did this using the {\sc jktabsdim} code \cite{Me++05aa} to utilise its distance-measurement capabilities. The results are given in Table~\ref{tab:absdim} and show that the masses are measured to precisions of 0.6--0.7\%, and the radii to 0.2--0.3\%. Using the \Teff\ values determined in the next section, we have calculated the luminosities and bolometric absolute magnitudes of the two stars. To our knowledge, this is the first published measurement of the radii of these stars.

The trigonometric parallax of \targ\ in \gaia\ DR3 is $15.367 \pm 0.022$~mas, a distance of $65.07 \pm 0.09$~pc, which allows a consistency check. We used the $BV$ and $JHK$ magnitudes in Table~\ref{tab:info}, the distance-determination method from Southworth et al.\ \cite{Me++05aa} and the surface brightness calibrations from Kervella et al.\ \cite{Kervella+04aa} to measure a distance of $63.67 \pm 0.76$~pc in the $K_s$ band. An interstellar extinction of $\EBV = 0.02 \pm 0.02$~mag was imposed to obtain consistent distance measurements in the optical and near-IR passbands. This extinction is appropriate for such a nearby object, and consistent with the $\EBV = 0.055 \pm 0.027$~mag given by the {\sc stilism}\footnote{\texttt{https://stilism.obspm.fr}} extinction maps \cite{Lallement+14aa,Lallement+18aa}.


\section*{Effective temperature from surface brightness ratio}

No published spectroscopic \Teff\ measurement exists for the individual components of \targ. G01 inferred spectral types of F8~V and G1-2~V, which correspond to \Teff s of approximately 6150~K and 5800~K in the calibration of Pecaut \& Mamajek \cite{PecautMamajek13apjs}. Two measurements exist from the Geneva-Copenhagen Survey \cite{Nordstrom+04aa,Casagrande+11aa} and are $\Teff = 6064 \pm 80$~K and 6030~K; both come from photometric calibrations based on Str\"omgren $uvby\beta$ indices and were obtained under the assumption that it is a single star. The \gaia\ DR3 {\sc gspphot} value is 6003~K \cite{Creevey+23aa} whilst the TESS Input Catalog \cite{Stassun+19aj} lists a slightly lower $5758 \pm 136$~K.

We therefore sought to obtain \Teff\ values for the two stars based on the \Teff\ of the system from Nordstr\"om et al.\ \cite{Nordstrom+04aa} and the known radius and surface brightness ratios from the {\sc jktebop} analysis. This is a straightforward procedure, but is not (to our knowledge) present in the literature so is outlined here. 

\begin{table} \centering
\caption{\em Physical properties of \targ\ defined using the nominal solar units given by 
IAU 2015 Resolution B3 (ref.\ \cite{Prsa+16aj}). \label{tab:absdim}}
\begin{tabular}{lr@{\,$\pm$\,}lr@{\,$\pm$\,}l}
{\em Parameter}        & \multicolumn{2}{c}{\em Star A} & \multicolumn{2}{c}{\em Star B}    \\[3pt]
Mass ratio   $M_{\rm B}/M_{\rm A}$          & \multicolumn{4}{c}{$1.1235 \pm 0.0045$}       \\
Semimajor axis of relative orbit (\Rsunnom) & \multicolumn{4}{c}{$49.24 \pm 0.10$}          \\
Mass (\Msunnom)                             &  1.0336 & 0.0059      &  1.1612 & 0.0081      \\
Radius (\Rsunnom)                           &  0.9787 & 0.0027      &  1.2105 & 0.0025      \\
Surface gravity ($\log$[cgs])               &  4.4711 & 0.0020      &  4.3370 & 0.0014      \\
Density ($\!\!$\rhosun)                     &  1.1024 & 0.0069      &  0.6546 & 0.0015      \\
Effective temperature (K)                   &  5890   & 100         &  6170   & 100         \\
Luminosity $\log(L/\Lsunnom)$               &   0.016 & 0.029       &  0.281  & 0.028       \\
$M_{\rm bol}$ (mag)                         &   4.699 & 0.074       &  4.036  & 0.071       \\
Interstellar reddening \EBV\ (mag)			& \multicolumn{4}{c}{$0.02 \pm 0.02$}			\\
Distance (pc)                               & \multicolumn{4}{c}{$64.20 \pm 0.80$}          \\[3pt]
\end{tabular}
\end{table}


First we make the assumption that the \Teff\ of the system ($T_{\rm sys}$) corresponds to the sum of the luminosities of the two stars ($L_{\rm A}$ and $L_{\rm B}$) so
\begin{eqnarray}
L_{\rm A} + L_{\rm B} & = & 4 \pi R_{\rm A} \sigma T_{\rm sys}^{~4} + 4 \pi R_{\rm B} \sigma T_{\rm sys}^{~4} \nonumber \\
                      & = & 4 \pi R_{\rm A} \sigma T_{\rm eff,A}^{~4} + 4 \pi R_{\rm B} \sigma T_{\rm eff,B}^{~4} ~,         \label{eq:lalb}
\end{eqnarray}
where $R_{\rm A}$ and $R_{\rm B}$ are the stellar radii and $\sigma$ is the Stefan-Boltzmann constant. Dividing by $4\pi\sigma$ and collecting terms gives
\begin{equation}
(R_{\rm A}^{~2} + R_{\rm B}^{~2}) T_{\rm sys}^{~4} ~=~ R_{\rm A}^{~2} T_{\rm eff,A}^{~4} + R_{\rm B}^{~2}T_{\rm eff,B}^{~4} ~.
\end{equation}
Replacing $R_{\rm B}$ with $kR_{\rm A}$ allows us to cancel out the radii:
\begin{eqnarray}
(R_{\rm A}^{~2} + k^2R_{\rm A}^{~2}) T_{\rm sys}^{~4} & = & R_{\rm A}^{~2} T_{\rm eff,A}^{~4} + k^2R_{\rm A}^{~2}T_{\rm eff,B}^{~4} \nonumber \\
(1+k^2) T_{\rm sys}^{~4} & = & T_{\rm eff,A}^{~4} + k^2T_{\rm eff,B}^{~4} ~.
\end{eqnarray}
Making the assumption that the radiative properties of the stars in the TESS passband are good proxies for their luminosities means that we can use the central surface brightness ratio, $J = T_{\rm eff,B}^{~4}/T_{\rm eff,A}^{~4}$ to get rid of $T_{\rm eff,B}$:
\begin{equation}
(1+k^2) T_{\rm sys}^{~4} ~=~ T_{\rm eff,A}^{~4} + k^2JT_{\rm eff,A}^{~4} ~=~ (1+k^2J)T_{\rm eff,A}^{~4} ~.
\end{equation}
We then rearrange to get the final result:
\begin{equation} \label{eq:TAsbr}
T_{\rm eff,A} ~=~ \left(\frac{1+k^2}{1+k^2J}\right)^{1/4} T_{\rm sys}
\end{equation}
after which we can obtain the \Teff\ of star~B from 
\begin{equation}
T_{\rm eff,B} ~=~ J^{1/4} \, T_{\rm eff,A} ~=~ \left(\frac{1+k^2}{\frac{1}{J}+k^2}\right)^{1/4} T_{\rm sys} ~.
\end{equation}
Due to the definition of $J$ in the {\sc jktebop} code, this formally requires the two stars to have the same limb darkening. However, the bias induced by this is small in general, and zero for \targ\ as the same limb darkening coefficients were used for both stars when fitting the light curve.

Following this procedure for \targ\ yields temperatures of $T_{\rm eff,A} = 5890$~K and $T_{\rm eff,B} = 6170$~K. The uncertainties in $k$, $J$ and $T_{\rm sys}$ were propagated using a Monte Carlo approach, and are dominated by that in $T_{\rm sys}$. We arbitrarily increased the uncertainties in $T_{\rm eff,A}$ and $T_{\rm eff,B}$ to 100~K to account for neglecting the wavelength dependence of $J$ in the above analysis. There will be an additional bias contributed by the assumption that the $T_{\rm sys}$ can be obtained from photometric indices of the combined system, but we lack the information necessary to quantify this (specifically the light ratios of the stars in the $uvby\beta$ passbands). The \Teff\ measurements presented here are simplistic, which means they are both limited and useful.


\section*{Effective temperature from light ratio}

The {\sc jktebop} code is well-suited to determining \Teff s via the central surface brightness ratio as this is one of its native parameters. However, some light curve models work instead with the light ratio so a different approach is needed to determine $T_{\rm eff,A}$ and $T_{\rm eff,B}$ from $T_{\rm sys}$. The equation is derived below for completeness, with the light ratio specified as $\ell = \ell_{\rm B}/\ell_{\rm A}$. Beginning with Eq.~\ref{eq:lalb} we can also write:
\begin{equation}
L_{\rm A} + L_{\rm B} ~=~ L_{\rm A}(1+\ell) ~=~ 4 \pi R_{\rm A} \sigma T_{\rm eff,A}^{~4} (1+\ell) ~.
\end{equation}
This step also requires the assumption that the measured light ratio in the observed passband is a good proxy for the luminosity ratio of the stars.

Substituting $R_{\rm B}$ with $k R_{\rm A}$, cancelling $4\pi\sigma R_{\rm A}$ as before, and then rearranging yields the final result:
\begin{equation} \label{eq:TAlr}
T_{\rm eff,A} ~=~ \left(\frac{1+k^2}{1+\ell}\right)^{1/4} T_{\rm sys} ~.
\end{equation}
The similarities between Eqs.\ \ref{eq:TAsbr} and \ref{eq:TAlr} are clear and are as expected from the physics of the situation. A similar approach but eliminating $L_{\rm A}$ instead of $L_{\rm B}$ gives the equation for the secondary star:
\begin{equation}
T_{\rm eff,B} ~=~ \left(\frac{1+\frac{1}{k^2}}{1+\frac{1}{\ell}}\right)^{1/4} T_{\rm sys} ~.
\end{equation}
For the record, this approach gave an identical result for \targ\ as the surface-brightness method above.


\section*{\targ\ in context}

The outstanding characteristic of \targ\ is, to us, the precise determination of the physical properties of two solar-type stars in an orbit of such a long period. In order to confirm and contextualise this, we sought comparable systems. For this we used the Detached Eclipsing Binary Catalogue \cite{Me15aspc} (DEBCat\footnote{\texttt{https://www.astro.keele.ac.uk/jkt/debcat/}}), which lists all known dEBs with mass and radius measurements to 2\% precision and accuracy. We required both components of a dEB to have a mass between 0.9 and 1.3\Msun\ and a surface gravity of $\log g > 4.0$ (c.g.s.), and the system to have a period of 10~d or more. A total of 14 dEBs (including \targ) satisfy the above criteria, of which \targ\ has the third-longest period. 

The dEBs are listed in Table~\ref{tab:debs} along with selected properties (mass, radius, period, eccentricity). Twelve of the 14 have a significant orbital eccentricity ($e > 0.16$), but any interpretation of this is complicated by the fact that eccentricity increases the likelihood of eclipses occurring \cite{Burke08apj,Winn10book}. Six of the dEBs were discovered in data obtained by space-based searches for transiting planets, and a further four have been studied using such data. Three of the 14 dEBs (Kepler-34, TIC 172900988 and Kepler-1647) have been studied in detail primarily because they host transiting circumbinary planets, and in these systems the presence of transits allows additional constraints on the properties of the inner binary system \cite{Doyle+11sci}. The list in Table~\ref{tab:debs} highlights the obvious advantage of extensive space-based photometry in the analysis of dEBs with long orbital periods and thus infrequent eclipses.

\begin{table} \centering
\caption{\em Identifications and properties of dEBs with similar properties 
to those of \targ, sorted in decreasing order of period. \label{tab:debs}}
\begin{tabular}{lccccccl}
Name & $P$ (d) & $e$ & $M_{\rm A}$\Msun & $R_{\rm A}$\Rsun & $M_{\rm B}$\Msun & $R_{\rm B}$\Rsun & Reference \\[3pt]
KX Cnc        & 31.220 & 0.470 & 1.134 & 1.053 & 1.124 & 1.059 & \cite{Me21obs1} \\
Kepler-34     & 27.796 & 0.521 & 1.048 & 1.162 & 1.021 & 1.093 & \cite{Welsh+12nat} \\
V454 Aur      & 27.020 & 0.381 & 1.034 & 0.979 & 1.161 & 1.211 & This work \\
KIC 7821010   & 24.238 & 0.680 & 1.277 & 1.276 & 1.221 & 1.210 & \cite{Helminiak+19mn} \\
LL Aqr        & 20.178 & 0.317 & 1.196 & 1.321 & 1.034 & 1.002 & \cite{Graczyk+16aa} \\
TIC 172900988 & 19.658 & 0.448 & 1.228 & 1.383 & 1.202 & 1.314 & \cite{Kostov+21aj} \\
V565 Lyr      & 18.799 & 0.020 & 0.995 & 1.101 & 0.929 & 0.971 & \cite{Brogaard+11aa} \\
LV Her        & 18.436 & 0.613 & 1.193 & 1.358 & 1.170 & 1.313 & \cite{Torres++09apj} \\
KIC 7177553   & 17.996 & 0.392 & 1.043 & 0.940 & 0.986 & 0.941 & \cite{Lehmann+16apj} \\
V963 Cen      & 15.269 & 0.422 & 1.081 & 1.445 & 1.075 & 1.421 & \cite{Graczyk+22aa} \\
AL Dor        & 14.905 & 0.195 & 1.102 & 1.092 & 1.103 & 1.098 & \cite{Graczyk+21aa} \\
Kepler-1647   & 11.259 & 0.159 & 1.221 & 1.790 & 0.968 & 0.966 & \cite{Kostov+16apj} \\
HP Dra        & 10.762 & 0.037 & 1.133 & 1.371 & 1.094 & 1.052 & \cite{Milone++10aj} \\
KIC 2306740   & 10.307 & 0.301 & 1.194 & 1.682 & 1.078 & 1.226 & \cite{Kocak+21apj} \\
\end{tabular}
\end{table}


\section*{Conclusion}

We have presented an analysis of the dEB \targ, which contains two solar-type stars on a relatively long-period ($P = 27.02$~d) and eccentric ($e=0.381$) orbit. We have determined the masses and radii of the component stars using light curves from eight sectors of TESS observations and extensive RVs obtained by G01. Our work provides the first published measurements of the radii of these stars. 

The system has the unusual characteristic that the star eclipsed at the deeper (primary) minimum is less massive, smaller \emph{and} cooler than its companion. This occurs because the stars are further apart during secondary minimum in this eccentric orbit, so a smaller fraction of the stars are eclipsed. More importantly, the physical properties are precisely determined and the stars are so far apart that tidal effects are negligible so they accurately represent the outcome of single-star evolution.

We used the measured temperature of the system plus the ratio of the radii and central surface brightnesses of the stars to determine their individual temperatures and thus luminosities. Our measured distance to the system is consistent with that from the \gaia\ DR3 parallax. A detailed study of the spectral characteristics of the stars could yield improved \Teff\ measurements as well as photospheric chemical abundances. \targ\ is therefore a promising candidate for conversion into a benchmark for the evolution of solar-type stars.

From a brief comparison of the masses, radii and \Teff s of the stars to the {\sc parsec} 1.2S theoretical stellar evolutionary models \cite{Bressan+12mn,Chen+14mn}, we find that the properties of the system are consistent with a solar chemical composition and an age in the region of $2.3 \pm 0.2$~Gyr.


\section*{Acknowledgements}

This paper includes data collected by the TESS\ mission and obtained from the MAST data archive at the Space Telescope Science Institute (STScI). Funding for the TESS\ mission is provided by the NASA's Science Mission Directorate. STScI is operated by the Association of Universities for Research in Astronomy, Inc., under NASA contract NAS 5–26555.
This work has made use of data from the European Space Agency (ESA) mission {\it Gaia}\footnote{\texttt{https://www.cosmos.esa.int/gaia}}, processed by the {\it Gaia} Data Processing and Analysis Consortium (DPAC\footnote{\texttt{https://www.cosmos.esa.int/web/gaia/dpac/consortium}}). Funding for the DPAC has been provided by national institutions, in particular the institutions participating in the {\it Gaia} Multilateral Agreement.
The following resources were used in the course of this work: the NASA Astrophysics Data System; the SIMBAD database operated at CDS, Strasbourg, France; and the ar$\chi$iv scientific paper preprint service operated by Cornell University.




\section*{Note added in proof}

\noindent After the completion and acceptance of the current work, an analysis of \targ\ was given by Y\"ucel, Canbay \& Baki\c{s} (arXiv:2404.18171). All parameters found by those authors agree with those found in the current work, representing a useful cross-check of our results. There were two significant differences. First, Y\"ucel et al.\ chose to identify the more massive star as the primary component. Second, the uncertainties in radius found by those authors are much larger (2--3\% versus our 0.2--0.3\%). The latter point is probably because the 120-s cadence data, and the data from sectors 71 to 73, were not available to Y\"ucel et al.\ at the time they began their analysis. Our results should be preferred as they are based on more extensive and better-sampled photometry.

\end{document}